\documentclass[a4paper,11pt]{article}
\usepackage{pos}

\title{Optimizing Staggered Multigrid for Exascale performance}
\ShortTitle{Multigrid for staggered fermions in 4D}

\author*[a]{Venkitesh Ayyar}
\author[a]{Richard Brower}
\author[b]{M.A. Clark}
\author[b]{Mathias Wagner}
\author[b]{Evan Weinberg}

\affiliation[a]{Boston University, Boston, MA 02215, USA}
\affiliation[b]{NVIDIA Corporation, Santa Clara, CA 95051, USA}

\emailAdd{vayyar@bu.edu}

\abstract{
Adaptive multi-grid methods have proven very successful in dealing with critical slow down for the Wilson-Dirac solver in lattice gauge theory. Multi-grid algorithms developed for Staggered fermions using the K\"ahler-Dirac preconditioning~\cite{Brower:2018ymy} have shown remarkable success. In this work, we discuss the performance of this staggered multi-grid algorithm in four dimensions. We also demonstrate that offloading some components of a multi-shift solve to a multi-grid solver leads to a significant performance improvement in an existing MILC spectrum workflow on the Summit and Selene supercomputers.
}

\FullConference{%
The 39th International Symposium on Lattice Field Theory,\\
8th-13th August, 2022,\\
Rheinische Friedrich-Wilhelms-Universität Bonn, Bonn, Germany
}

\graphicspath{{./}{figures/}}
\usepackage{subfiles}
\usepackage{subfig}
\usepackage{amsmath}
\usepackage{multicol}

\begin{document}
\maketitle

\section{Introduction}

The era of exascale computing has finally arrived. After years of improvement and innovation on the hardware level, as well as innovations in software and algorithms which more optimally utilize such hardware, we are now capable of running calculations that exceed one exaflop of computing throughput. This accomplishment is indicative of a commensurate boost in all computing workflows, and by extension a growth in scientific challenges we can pursue with a computational approach.

Unfortunately, raw flops do not alone unlock super-linear growth in the number of problems we can tackle. Cutting-edge problems of interest do not scale trivially with the increase in computing power. The path towards more precise computational science suffers from a phenomena known as critical slowing down~\cite{Blum:2013mhx}, where the cost of meaningful computation scales with a higher exponent than the na\"ive arithmetic complexity of the algorithms in use.

One of the clearest indicators of critical slowing down in lattice gauge theory simulations is non-trivial increases in the costs of one key kernel in most workflows: the iterative solution of the Dirac linear equation as part of propagator calculations and (rational) hybrid Monte Carlo (RHMC) evolution~\cite{Duane:1987de,Clark_2007}. In the approach to the continuum limit, the number of iterations required to solve the Dirac linear system to a fixed tolerance grows super-linearly. This is best understood in terms of the non-linear increase in the condition number of the system as the lattice spacing $a$ decreases.


There are many methods which stave off but do not solve critical slowing down. One is block-Krylov solvers, which improve reuse across multiples solves but do improve the condition number of the matrix~\cite{Nikishin:1995,Clark:2017ekr}. A second is eigenvalue deflation~\cite{Lanczos1950AnIM} and improvements thereof in~\cite{de_Forcrand_1996,Morgan2007,Stathopoulos_2010,Clark:2017wom,ROMERO2020109356}, to name a few, which improves the condition number of the matrix by performing an exact solve of the low space. This approach thus shifts the issue of critical slowing down to the eigensolve. The one approach that solves critical slowing down in a scalable fashion is a multi-grid (MG) algorithm.

Adaptive multi-grid preconditioning with geometric aggregation have shown remarkable success in mitigating critical slowing down in lattice gauge theory applications. The first success was with the Wilson and Wilson-clover formulations~\cite{Babich:2010qb,Osborn1011,Brannick:2007ue} (complimented by the similar algorithm of inexact deflation in~\cite{L_scher_200_inexact}). There have also been successful extensions to twisted-mass and -clover fermions~\cite{Frezzotti:2000nk,doi:10.1137/130919507,Richtmann_2022}, as well as chiral domain wall fermions~\cite{Brower:2020xmc,Cohen:2012sh,Boyle:2014rwa}.

A remaining challenge is the development and deployment of an MG algorithm for Kogut-Susskind, or staggered, fermions~\cite{PhysRevD.11.395}. The mathematical framework of a staggered MG algorithm was developed in two dimensions in~\cite{Brower:2018ymy}. In this work, we describe the extension and implementation of an MG algorithm for staggered fermions in four dimensions. We describe the details of the implementation and share performance on the Summit Supercomputer at Oak Ridge National Laboratory~\cite{top500_summit}, as well as the Selene Supercomputer at NVIDIA~\cite{top500_selene}. Last, we share thoughts on future directions in the development and optimal implementation of multi-grid solvers.

\section{Multi-grid methods for Staggered fermions}
\subsection{Multi-grid algorithms}\label{mg}

\begin{figure}[ht!]
\includegraphics[width=0.85\textwidth]{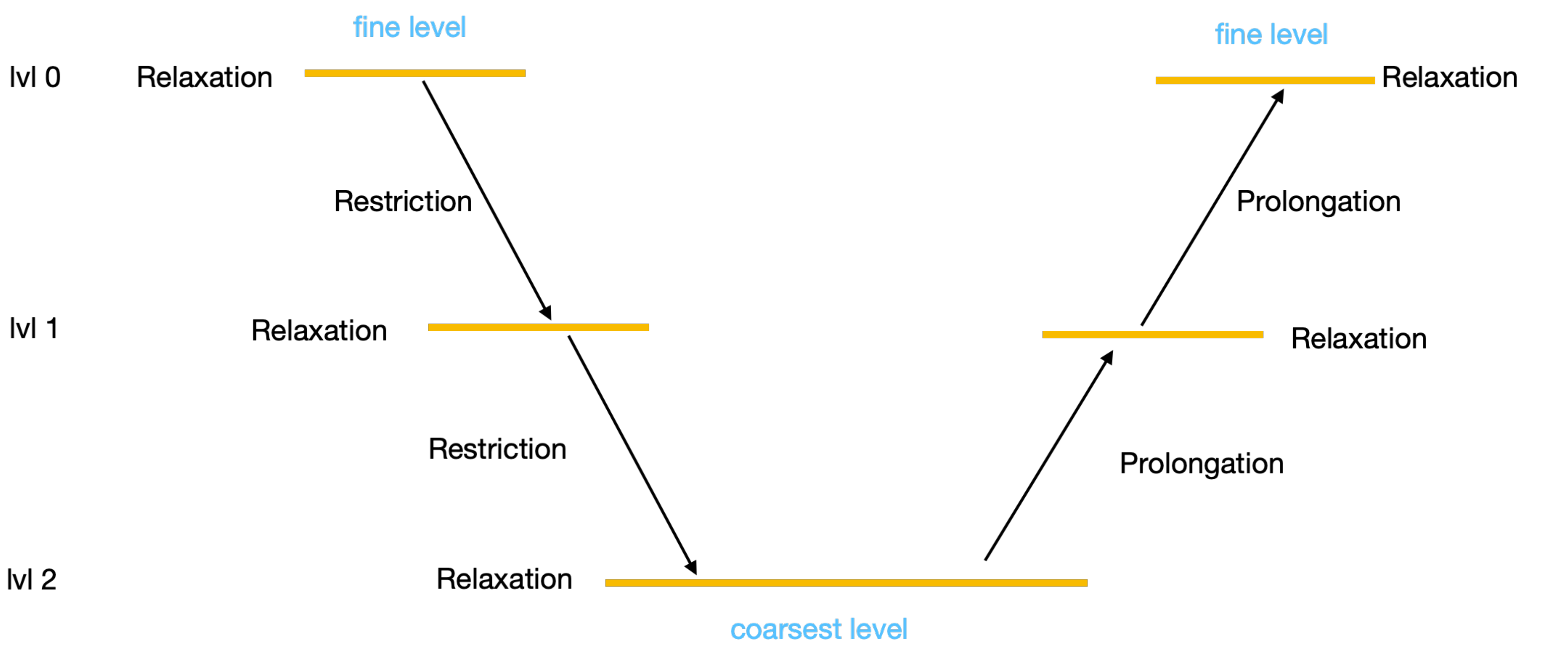}
\caption{Schematic of a typical MG algorithm with three levels. At each level, the residue after the relaxation operation is restricted to the coarser level, where it serves as the source for further relaxation operations. The prolongation operation converts the solution at the coarser level to an error correction at the finer level, which is added to the current accumulated solution.}
\label{fig:mg_schematic}
\end{figure}

At a high level, the idea of an MG approach is to solve, in tandem, modes at all scales in a given Dirac operator. We solve this as a \textit{K-cycle}, where the MG preconditioner is the preconditioner in an outer (flexible) Krylov solver; here we use generalized conjugate residual (GCR)~\cite{Axelsson1987}. The preconditioner is implemented as follows: (1) pre-smooth the current residual, (2) ``restrict,'' or aggregate, the current residual with a restriction operator to the coarser level, (3), iterate on the coarser level linear system to some fixed tolerance, (4), ``prolongate'' the error correction from the coarser level to the fine level, updating the solution, and (5) post-smoothing the new residual. A schematic of this is given in Fig.~\ref{fig:mg_schematic}. This is extended to a recursive algorithm by applying an MG preconditioner to the iterative solve on the coarser level. This is described in greater detail in~\cite{Brower:2018ymy}.

The restriction operator $R$ is formed from block-orthonormalized \textit{near-null vectors}, which are vectors rich in low modes. Here, near-null vectors are generated via inverse iterations on the \textit{homogeneous system}, $A \vec{\phi}_0 = \vec{0}$, where $\vec{\phi}_0$ is seeded with random numbers. After a large number of iterations the solution vector will be rich in low modes. We perform a chiral doubling on each near-null vector which preserves a form of ``$\gamma_5$''-Hermiticity on the coarse level; the staggered chiral projector is defined in the following subsection. The process of block-orthonormalization, where the block size defines the aggregation factor, increases the span of the fine space which is preserved on a coarser level.

Given a restriction operator $R$, we define the projection operator $P$ via a Galerkin prescription, $P = R^\dagger$. The coarse operator is explicitly constructed as $\hat{A} = R A P$ where the $\hat{\cdot}$ notation denotes the ``coarsened'' version.

\subsection{Staggered fermion and K\"ahler-Dirac preconditioning fundamentals}

The lattice staggered fermion formulation~\cite{PhysRevD.11.395} trades some residual fermion doubling for an exact lattice chiral symmetry and is equivalent to the K\"ahler-Dirac formulation~\cite{PhysRevD.38.1206} in the free field, where there is a one-to-one equivalence between the $2^d$ hypercube of degrees of freedom for staggered fermions and the $2^d$ degrees of freedom for K\"ahler-Dirac fermions. There is an exact chiral symmetry generated by $\gamma_5 \otimes \tau_5 = \epsilon(x) = \left(-1\right)^{x + y + ...}$, where $\gamma$ and $\tau$ denote the spin and taste space, respectively, and by extension the chiral projector is defined by $\frac{1}{2}\left(1 \pm \epsilon(x)\right)$. The eigenspectrum of the staggered operator is maximally anti-Hermitian indefinite up to a real mass shift.

The benefits of the physics of the staggered fermion turns into a challenge for staggered fermion MG. In four dimensions, the four-fold increase in the number of degrees of freedom of a staggered fermion relative to a Wilson fermion increases the size of the necessary near-null space, with a commensurate increase in computational complexity. More fundamental is that the general adaptive MG algorithm described above, which is successful for the Wilson fermion formulation, fails due to \textit{spurious small eigenvalues} in the coarsened staggered operator. 


\begin{figure}[t]
\centering
\subfloat[\centering Representative spectrum of the free-field staggered and KD-preconditioned operator in two dimensions~\cite{Brower:2018ymy}.]{{\includegraphics[width=0.6\textwidth]{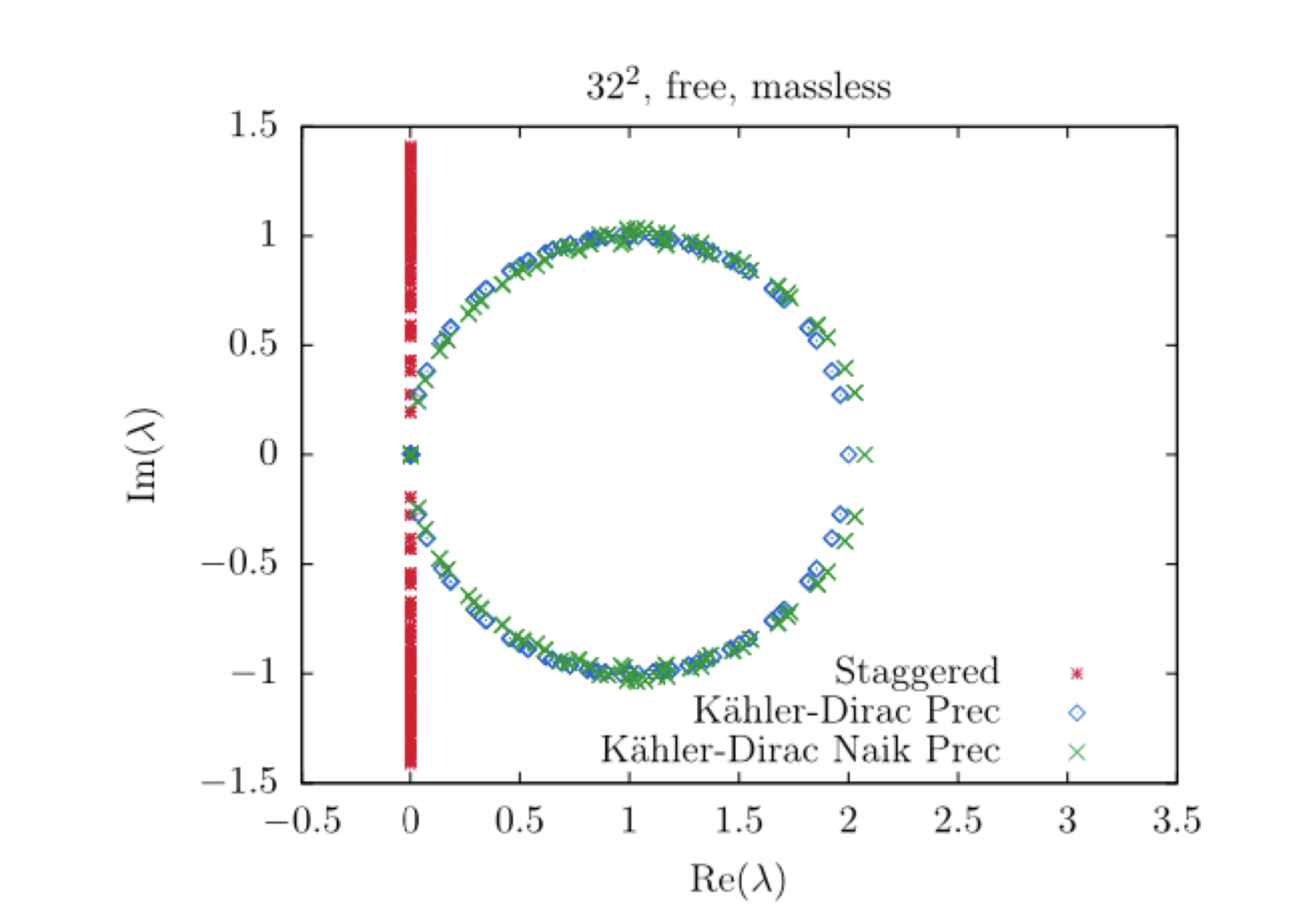}}}
\qquad
\subfloat[\centering General schematic of the staggered MG algorithm.]{{\includegraphics[width=0.3\textwidth]{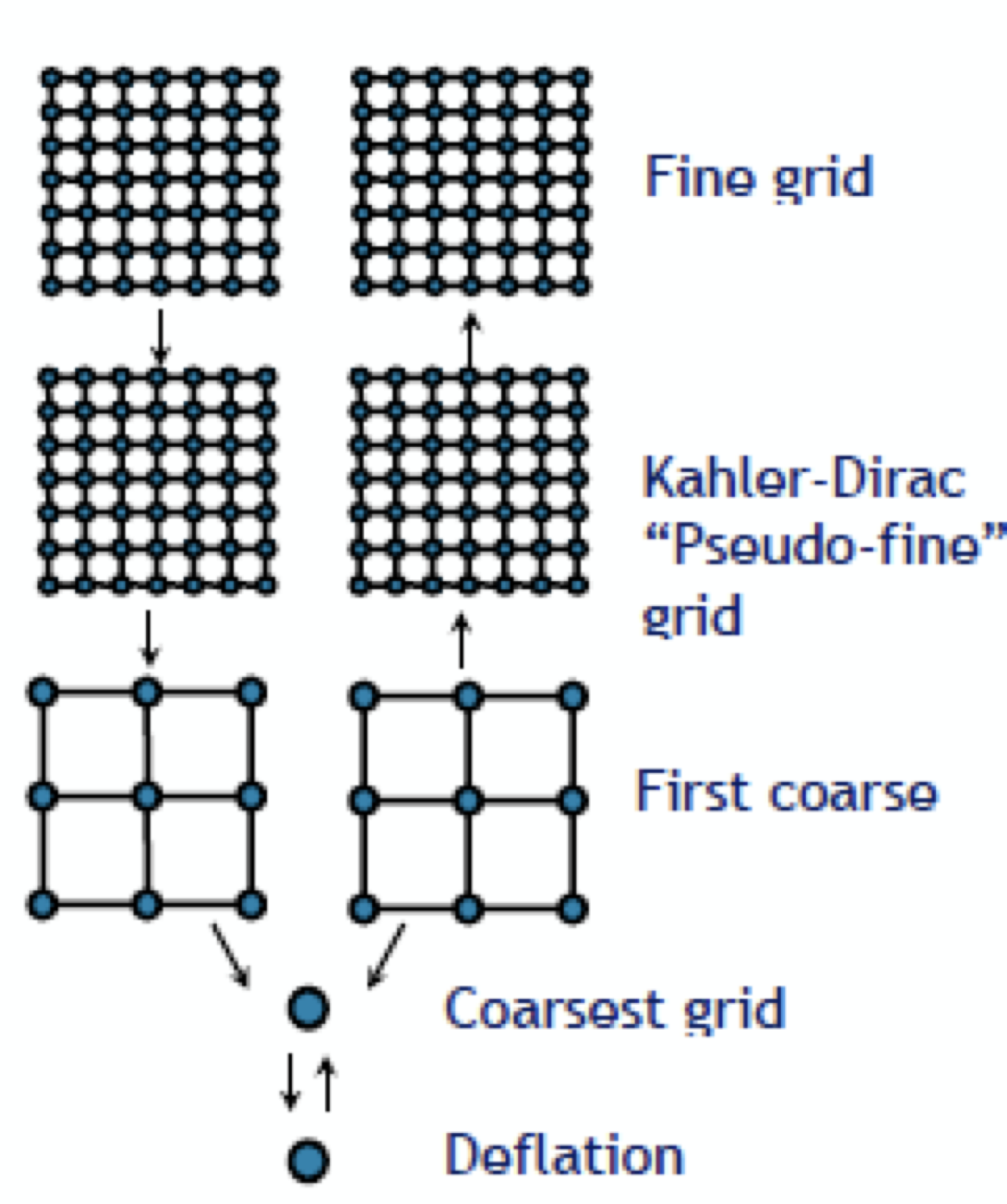}}}
\caption{On the left, a comparison of the spectrum of the KD-preconditioned and non-preconditioned staggered operator. The application of the KD preconditioner converts the staggered spectrum into an overlap-esque spectrum. On the right, a visual representation of the structure of the five-level staggered-type MG algorithm. Of note, the second level is the KD-preconditioned operator which still acts on the fine space, and the fifth level is a deflation space.}
\label{fig:comparison_schematic}
\end{figure}

It was found in~\cite{Brower:2018ymy} that there is a solution to this problem: a so-called K\"ahler-Dirac (KD) preconditioner, which deforms the eigenspectrum of the staggered operator to a form more amendable to MG preconditioning. We consider the unimproved staggered Dirac operator as a sum of two contributions: the hopping terms that stay \textit{within} the $2^d$ hypercube, denoted $B$, and the hopping terms that \textit{connect} $2^d$ hypercubes, denoted $C$. In the language of the equivalent K\"ahler-Dirac operator, these are the internal degrees of freedom as opposed to the matrix elements connecting sites. The operator can be written as:
\begin{equation}
    D_{\mbox{stag}}= \left(B + m\right) + C,
\end{equation}
which constitutes a \emph{dual-decomposition}. \textit{In the free field}, up to a scaling, the terms $\left(B + m\right)$ and $C$ are each anti-Hermitian (up to a real mass shift). Each obeys an identical $\epsilon$-Hermiticity. $B^\dagger B$ and $C^\dagger C$ are proportional to the identity matrix, implying they are each separately {\emph{unitary}}.

The matrix $B$ is block-local: in the free-field case, each hypercube contains $2^d$ independent sites, or $2^d$ degrees of freedom. This makes taking the inverse of $B$ trivial. Kahler-Dirac preconditioning involves left- (or right-)multiplying $D_{\mbox{stag}} $ with $B^{-1}$ to obtain 
\begin{equation}
    A = B^{-1} D_{\mbox{stag}} = \mathcal{I} + B^{-1} C = \mathcal{I} + \left(\epsilon B\right)^{-1} \left (\epsilon C\right).
\end{equation}
Since $\epsilon B$ and $\epsilon C$ are each unitary in the free field, their product is also unitary, and thus the final K\"ahler-Dirac preconditioned operator has a shifted-circular, a.k.a. overlap-esque, structure. This is shown on the left-hand side of Fig.~\ref{fig:comparison_schematic}. We perform an MG aggregation on this operator.

The observations we make here break down in the interacting case as neither operator is unitary. Further, in the four-dimensional case, the more relevant operator is the HISQ~\cite{Orginos:1999cr,Follana:2006rc} operator. Here the dual decomposition breaks down due to the introduction of a Naik term. However, everything noted above remains \textit{approximately} true in these cases, so we put our blinders on and carry on with the KD-preconditioned operator.

\subsection{K\"ahler-Dirac preconditioned multigrid}

We now describe the high-level implementation of our algorithm. A complementary sketch of the process is given in the right-hand side of Fig.~\ref{fig:comparison_schematic}. First, our outer-most operator is the full HISQ stencil containing the fat and long links. The next recursive level is the K\"ahler-Dirac preconditioned operator. This is an abuse of language since this next ``level'' does not include a thinning of degrees of freedom. We nonetheless insert it into the same algorithmic structure---including a pre- and post-smoother surrounding it---and denote this level as ``pseudo-fine'' to reflect its nature.

Subsequent levels are constructed as expected: generate near-null vectors, perform a chiral doubling by applying $\frac{1}{2}\left(1 \pm \epsilon(x)\right)$, perform a block orthonormalization, construct a prolongator, restrictor, and coarse operator, then recurse. As described in previous publications, we take the non-traditional approach of generating near-null vectors with the coarse operator and then coarsening the \textit{block-preconditioned} version of said operator. In general this \textit{improves} the condition number on each level, analogous to even-odd preconditioning for the Wilson-clover operator~\cite{DeGrand:1990dk}.

As a final optimization, we perform a deflation with the low-lying singular values on the coarsest level. This is inspired by deflation of a Hermitian positive definite operator in~\cite{https://doi.org/10.48550/arxiv.1611.06944}, however, the key innovation in the QUDA library is using singular value decomposition deflation instead. The use of deflation in the workflow is an acknowledgement that recursing to an arbitrarily small problem size can be inefficient. This means that we do \textit{not} have a perfect MG algorithm, and there is residual critical slowing down on the coarsest level. The deflation addresses this, with the benefit that because it is a much smaller operator, generating the eigenspace is more efficient than generating the eigenspace for the fine operator.


\section{Implementation and Results}
\subsection{Details of the implementation}

For our investigations here, we utilize the efficient implementation of MG workflows in the QUDA library for GPUs~\cite{Clark:2009wm,Babich:2010mu,Babich:2011np,Clark:2016rdz}. Originally developed for NVIDIA GPUs, it now features a performance-portable abstraction that includes formal support for HIP to target AMD GPUs, as well as near-complete support for SYCL and work-in-progress support for OpenMP device targets.

QUDA's solvers utilize reduced-precision methods, as well as gauge-link compression based on their inherent symmetries. HISQ long links, being in the $U(3)$ group, can be compressed as 9 or 13 real numbers. Conjugate Gradient (CG) solvers run in mixed precision, here mixed double/half, where ``half'' refers to a 16-bit fixed point format with a per-site fp32 norm. The staggered and HISQ stencil implementation features GPU-initiated communications via NVSHMEM support when run on NVIDIA GPUs.

The coarse stencil operator is also highly optimized; the coarse links are stored in 16-bit fixed point format, and additional parallelism is exposed in the matrix-vector application to compensate for reduction in parallelism possible by the size of the local volume alone. Halo exchanges for the coarse spinor fields utilize NVSHMEM packing kernels on NVIDIA GPUs. This is an important optimization given that communications on the coarse level are inherently latency limited.


Our implementation of the KD-preconditioned operator has evolved significantly over the course of this work. We initially performed a unitary transformation of the staggered operator, acting on all $V$ sites on the lattice with $N_c = 3$ degrees of freedom per site, into an operator acting on $\frac{V}{2^d}$ ``super-sites'' with $2^d$ times $N_c = 3 \rightarrow 48$ degrees of freedom. The stencil of this operator included many zeroes in deterministic locations. Next, we constructed the inverse of the block-local term $B$ and pre-applied it to each ``hopping'' term. In the HISQ case, distance-three Naik terms on the KD-preconditioned operator were dropped. This was justified via a perturbative argument: the leading coefficient of the Naik term is $5\%$ of the magnitude of the distance one fat link terms. This implementation required storing nine $48^2$ matrices per $2^d$ super-site: eight for the pre-computed hopping terms and one for the KD inverse term $(B+m)^{-1}$. This corresponds to $9 \times 48^2 / 16 = 1296$ complex numbers per fine site.

The revised approach is to apply the KD operator in a form that takes the decomposition $A = (B+m)^{-1} D_{\mbox{stag}}$ literally. First, we apply the HISQ operator, taking full advantage of the optimizations already present in the QUDA. Next, we apply the local KD operator in a separate kernel. It is a point of future optimization to fuse these operations.

In contrast to the initial brute-force implementation, the optimized operator requires loading the existing fat and long $3 \times 3$ links in four directions per fine site on the $V$-sized lattice, along with the same explicit $(B+m)^{-1}$ matrix per $\frac{V}{2^d}$ sites. This corresponds to $8 \times 3^2 + 48^2 / 16 = 180$ complex numbers per fine site: a 72\% reduction. An extra benefit is the fine links do not require additional storage because they can be reused from the outer HISQ operator.

One benefit to this formulation is we have the flexibility to preserve or drop the Naik contribution. When we preserve the Naik term, the KD-preconditioned operator is an exact left preconditioning of the full HISQ operator. When we truncate the Naik contribution, it becomes a less effective preconditioner on paper, however it shows comparable, if not better, performance during runs at scale due to a factor-of-two reduction in memory traffic during the stencil application and a factor-of-three reduction in data communicated.

In the following subsection we present results from solving the HISQ linear system with an MG preconditioner. We only present results for the optimized implementation of the full HISQ operator and the truncated operator because the memory overheads of the brute force operator lead to at least a doubling of the number of required nodes.

\subsection{Performance comparison of Multigrid vs Conjugate Gradient} \label{base_comparison}

While the algebraic description above suggests a successful MG algorithm, it does not prove the viability of the technique in production workflows. The one true metric of success is an improvement in \textit{time to solution} for propagator solves.

For our studies here, we consider a $144^3 \times 288$, 2+1+1 physical pion mass HISQ configuration shared by the MILC collaboration. The configuration is very fine, with a lattice spacing of 0.04 fm, and has a bare light and strange quark mass of 0.000569 and 0.01555, respectively.

We distribute the calculation over 864 GPUs. We perform a $6 \times 3 \times 6 \times 8$-way partitioning of the global volume, leading to a per-GPU local volume of $24 \times 48 \times 24 \times 36$. For the Summit supercomputer, with 6 NVIDIA V100 GPUs per node, this corresponds to 144 nodes. By ordering the decomposition $x$-fastest, this selection reflects the organization of GPUs on nodes and nodes within racks on Summit: with 6 GPUs per node and 18 nodes per rack, all $x$ communications stay within the node, $y$ and $z$ stay within the rack, and $t$ goes inter-rack. On the other hand, for Selene, with 8 NVIDIA A100 GPUs per node, corresponding to 108 nodes, we order the decomposition $t$-fastest so all $t$ communications stay within the node.

Many parameters go into tuning an MG solve between the near-null space generation, coarsest-level SVD deflation, and the parameters of the MG solve itself. We summarize the parameters of our HISQ workflow in Table~\ref{tab:mg_setup}. We note that for the smoother and for the coarsest-level solve we have developed a communication-avoiding s-step version of GCR, denoted CA-GCR, an extension of the idea developed in~\cite{ChronoGMRES} for generalized minimum residual (GMRES).

\begin{table}[t!]
\tiny\centering
\begin{tabular}{|r|c|l|}
\hline
 Level & Property & Value \\
\hline\hline
 0 & Solver & MG-preconditioned GCR \\
 \hline
 0 & Operator & Full (non-Schur) HISQ operator \\
 \hline
 0 & Degrees of freedom & $N_c = 3$ \\
 \hline
 0 & Pre-, post-smoother & none, CA-GCR(8) \\
 \hline\hline
 0 and 1 & Per-rank volume & 24 x 48 x 24 x 36 \\
 \hline\hline
 1 & Solver & MG-preconditioned GCR \\
\hline
 1 & Operator & KD-preconditioned HISQ {\bf or} truncated operator \\
 \hline
 1 & Degrees of freedom & $N_c = 3$ \\
 \hline
 1 & Solver convergence & 0.25 {\bf or} 8 iterations\\
 \hline
 1 & Pre-, post-smoother & none, CA-GCR(8) \\
 \hline
 \hline
 1 $\rightarrow$ 2 & Setup solver & CGNE \\
 \hline
 1 $\rightarrow$ 2 & Setup tolerance & $10^{-6}$ {\bf or} 2000 iterations \\
 \hline
 1 $\rightarrow$ 2 & Aggregate size & 4 x 6 x 6 x 6 \\
 \hline\hline
 2 & Solver & MG-preconditioned GCR \\
 \hline
 2 & Operator & Schur-preconditioned \\
 \hline
 2 & Degrees of freedom & $N_c = 64, N_s = 2$ \\
 \hline
 2 & Solver tolerance & 0.25 {\bf or} 8 iterations \\
 \hline
 2 & Pre-, post-smoother & none, CA-GCR(8) \\
 \hline
 \hline
 2 $\rightarrow$ 3 & Setup solver & CGNE \\
 \hline
 2 $\rightarrow$ 3 & Setup tolerance & $10^{-6}$ {\bf or} 2000 iterations \\
 \hline
 2 $\rightarrow$ 3 & Aggregate size & 3 x 2 x 2 x 3 \\
 \hline
 \hline
 3 & Solver & SVD-deflated Chebyshev-basis CA-GCR(16) \\
 \hline
 3 & Operator & Schur-preconditioned \\
 \hline
 3 & Degrees of freedom & $N_c = 96, N_s = 2$ \\
 \hline
 3 & Singular vectors & 1024 \\
 \hline
 3 & Deflation acceleration & Yes, 400-degree Chebyshev polynomial \\
 \hline
\end{tabular}
\caption{\label{tab:mg_setup}Key parameters for the MG preconditioner.}
\end{table}

\begin{figure}
\centering
\subfloat[\centering Comparison of timings for MG and CG runs]{{\includegraphics[width=0.45\textwidth]{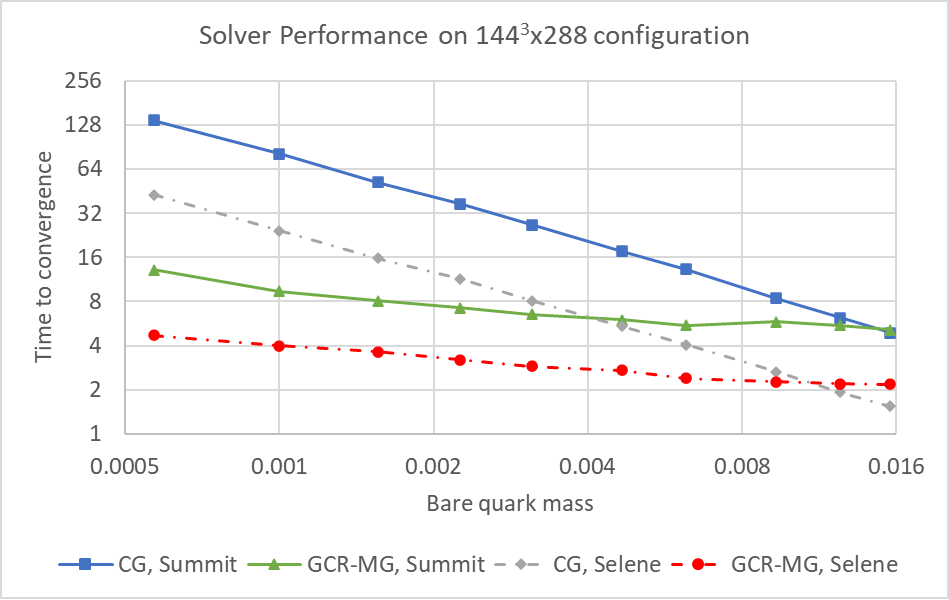}}}
\qquad
\subfloat[\centering Ratio of timings from MG to CG on Summit]{{\includegraphics[width=0.45\textwidth]{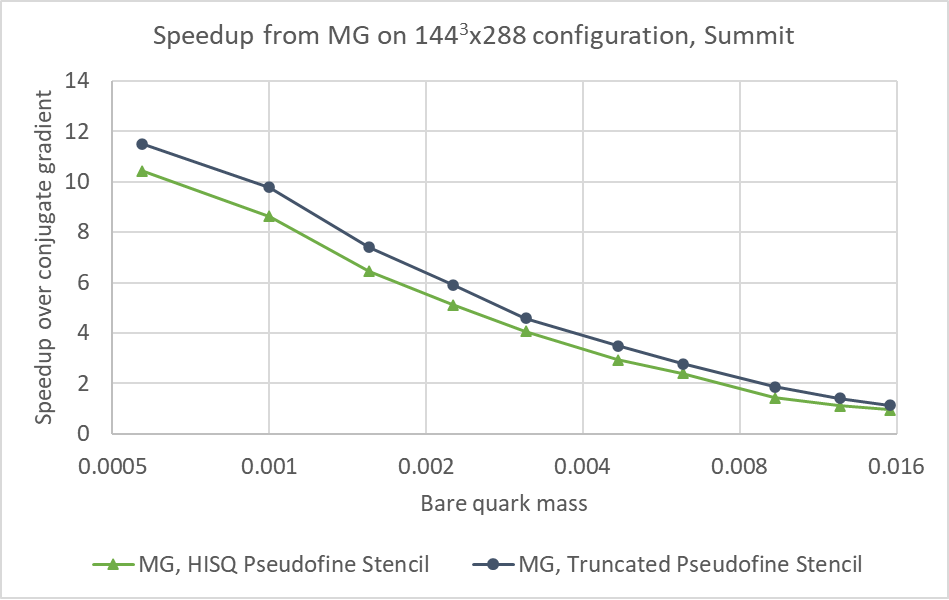}}}
\caption{On the left, the time to solution for conjugate gradient and MG-preconditioned GCR on Summit (solid lines) and Selene (dashed lines). The MG-preconditioned solve uses the full HISQ stencil in the pseudo-fine operator. On the right, the relative performance of MG-preconditioned solves---with the full HISQ stencil and the truncated stencil in the pseudo-fine operator---to a standalone CG solve on Summit.}
\label{fig:mg_vs_cg}
\end{figure}

In Fig.~\ref{fig:mg_vs_cg}, we show the solver performance of our staggered MG algorithm compared with the typical Schur-preconditioned CG solve for the HISQ stencil. In all cases we performed the solve to a relative tolerance of $10^{-10}$ on the full (non-Schur) operator. In the case of the single-parity CG solve, this requires a solve to the relative tolerance $m \times 10^{-10}$.

On the left, we show the performance of CG and MG-preconditioned GCR on both the Summit and Selene supercomputers. In this case, we are using the full HISQ stencil in the KD-preconditioned operator, that is, we do not drop the Naik term. While the cost of the CG solve scales with the inverse of the mass, corresponding to the expected scaling of the condition number, we see that the MG-preconditioned solve shows only a slight dependence on the mass. This is an indication of successful removal of critical slowing down.

On the right, we show the relative performance of our MG algorithm against a CG solve on the Summit supercomputer. Here, we present two curves: the performance boost for the full HISQ stencil in the KD-preconditioned operator (green triangles), and the performance boost for the \textit{truncated} stencil in the KD-preconditioned operator (black circles). For the smallest mass, $m = 0.000569$, the performance gain is more than 10x for both the full and the truncated operator. Perhaps even more impressively, there is rough performance parity at the strange quark mass, $m = 0.01555$.

There is an important note to go along with this data. We generate the near-null vectors and singular vectors on the coarsest level \textit{once} using the operators formed from the smallest mass, $am = 0.000569$. We save these and reuse them for each subsequent solve. This approach is an emulation of a proper propagator workflow, where the costs of re-generating near-null vectors and singular vectors, along with re-formulating the new coarse operator, could become end-to-end prohibitive.

We note that these numbers do not include the setup costs, these are discussed in the context of a real workflow in the following subsection.

\subsection{Multigrid performance on a MILC workflow on Summit}

It is clear from section \ref{base_comparison} that the staggered MG algorithm in four dimensions offers a very significant benefit over CG for light masses. However, it is important to understand how useful this algorithm in actual lattice gauge theory workflows with multiple masses and greater complexity. In this section, we show such a comparison of MG and CG for a model pion/kaon propagator workflow.

The workflow at hand involves computing propagators for ten different quark masses, ranging from the physical light quark to the strange quark. Typically this workflow is run using {\it multi-shift CG}, where the propagator for each mass shift is solved simultaneously with the overhead of only one matrix-vector application per iteration.

In the most general case it is challenging to formulate a preconditioned multi-shift CG. This makes applying the MG preconditioner described here non-trivial. The simplest, and effective, solution is to apply an MG preconditioner to the smallest subset of masses, where there is the greatest benefit~\cite{Alexandrou_2019}. We reuse the coarse operators from the lightest mass for other masses in this subset. The remaining masses, equivalently shifts, are still solved in what is now a much less expensive multishift CG. For this workflow, we split the ten masses 3:7 between MG and multi-shift, an empirical decision that will be better explored in the future.

Table~\ref{tab:comparison1} gives a comparison of normalized timings for MG and CG on the Summit supercomputer, where we have used 144 nodes as described above for MG, and 72 nodes for CG, owing to the reduced memory overheads. The propagator time corresponds to the time it takes to solve for three right hand sides for an MG solve, and six for CG solves---three for the even system, three for the odd. We see that, when normalized by node-hours, the cost of MG setup is amortized after roughly three propagator calculations.

\begin{table}
\centering
\begin{tabular}{|r|c|c|c|} 
\hline
 Run type & Summit Nodes & MG Setup & One Propagator \\ 
 \hline
 Units & --- & Node-hours & Node-hours \\ 
 \hline\hline
 Multishift CG & 72 & --- & $9.87 \times 10^4$ \\ 
 \hline
 MG+CG (full op) & 144 & $2.28 \times 10^5$ & $3.34 \times 10^4$ \\ 
 \hline
 MG+CG (truncated op) & 144 & $1.89 \times 10^5$ & $3.37 \times 10^4$ \\ 
\hline
\end{tabular}
\caption{\label{tab:comparison1}Performance comparison of mixed MG and multishift CG \textbf{vs} just multishift CG for a propagator workflow with 10 masses ranging from the light to the strange quark. We note that the break-even point for the full pseudo-fine operator is $\approx 3.5$ propagators, while for the truncated pseudo-fine operator it is $\approx 2.9$ propagators. Run on Summit.}
\end{table}
\section{Conclusion}
\subsection{Summary}

In this work we have described a successful implementation of a multi-grid solver for the HISQ operator in four dimensions. More importantly, we have demonstrated the success of the algorithm on a very fine $144^3 \times 288$ physical pion mass HISQ configuration, showing the near-elimination of critical slowing down across a range of masses from the physical light up to the strange quark mass. Using the QUDA library for lattice calculations on GPUs, we have demonstrated a 10x speedup in solver time to a fixed tolerance at the light quark mass for a solve distributed over 144 nodes of the Summit supercomputer, and shown we can break even in solve time at the strange quark mass.

Due to the multi-year effort of performance portability work that went into the QUDA library, we have also successfully run this algorithm on a smaller configuration on the Crusher supercomputing testbed at ORNL.

\subsection{Future directions}

A key challenge of the current algorithm is the high overhead of generating several near-null vectors for the HISQ operator due to the large number of low modes inherent to the staggered formulation. We are currently exploring multiple avenues to improve the setup time. From an algorithmic standpoint, this includes, but is not limited to, multi-right hand side solver methods, Chebyshev filter approaches to generating near-null vectors~\cite{Boyle2103}, and block-TRLM methods for singular vector generation.

Although this algorithm has a clear use case in the measurement of correlation functions in lattice gauge theories, there is potential for using it in the gauge generation process as well. One challenge in using the MG solver in HMC is that we must evolve the near-null vectors as the gauge field evolves. This has been successfully applied in the Wilson-clover case~\cite{L_scher_2007} and the Shamir domain wall case~\cite{Boyle2103}, however the relatively higher setup costs for HISQ MG and the presence of multi-shift solves in RHMC make this less trivial. These studies will be a follow-up to the improved setup investigations above.

\begin{acknowledgments}
This work was supported in part by the U.S. Department
of Energy (DOE) under Award No. DE-SC0015845 and by the Exascale Computing Project (17-SC-20-SC), 
a collaborative effort of the U.S. Department of Energy Office of Science and the National Nuclear Security Administration. For computation, we used the GPU nodes of the Summit and Crusher supercomputers at Oak Ridge National Laboratory.
\end{acknowledgments}

\bibliography{main}{}
\bibliographystyle{siam}

\end{document}